\documentclass[aps,prl,reprint,superscriptaddress,floatfix,nofootinbib,showkeys,showpacs,preprintnumbers]{revtex4-2}
\pdfoutput=1

\usepackage{tabularx}
\usepackage{xspace}
\usepackage{listings}
\usepackage{lineno}
\usepackage{bm}		%
\usepackage{amsmath}	%
\usepackage{amssymb}	%

\usepackage[dvipsnames,svgnames]{xcolor} 
\usepackage{graphicx}
\usepackage{comment}

\usepackage{natbib}
\usepackage{hyperref}
\usepackage{amsmath}
\usepackage{enumitem}
\setlist{wide, labelwidth=!, labelindent=0pt}
\usepackage{multirow}
\hypersetup{    
  colorlinks      = true,
  linkcolor       = {black},
  linkbordercolor = {white},
  citecolor       = {black},
  citebordercolor = {white},
  urlcolor        = {blue},
  urlbordercolor  = {white},
}
\usepackage{soul} 
\usepackage{amsmath}
\usepackage{amssymb}
\usepackage{xspace}
\usepackage{xifthen}

\mathchardef\mhyphen="2D

\newlength{\dhatheight}

\newcommand{\bandvar}[2][]{%
  \ifthenelse{\isempty{#1}}{\var{#2}}{\var{#2\_#1}}%
}

\newcommand{\var}[1]{\ensuremath{\texttt{\MakeUppercase{#1}}}\xspace}

\newcommand{\redmagic}{\rm redMaGiC}

\newcommand{\ccz}{cross-correlation}

\newcommand{\cczs}{cross-correlations}

\newcommand{\omegam}{\Omega_{\rm{m}}}
\newcommand{\sigmae}{\sigma_8}

\newcommand{\clustercomb}{4$\times$2pt+N}
\newcommand{\allcomb}{6$\times$2pt+N}
\newcommand{\ttt}{3$\times$2pt}

\defcitealias{Simpaper}{Simpaper}
\defcitealias{DESY1KP}{DESY1KP}
\defcitealias{DES_cluster_cosmology}{DESCL}

\providecommand\physrep{\ref@jnl{Phys.~Rep.}}%
\providecommand\apjs{\ref@jnl{ApJS}}%
\providecommand{\jcap}{\ref@jnl{JCAP}}%
 \usepackage{lipsum}
\usepackage{aas_macros}
\newcommand{\redmapper}{{\rm redMaPPer}}

\begin{document}
\preprint{DES-2020-579}
\preprint{FERMILAB-PUB-20-465-AE}

\title{Dark Energy Survey Year 1 Results: Cosmological Constraints from Cluster Abundances, Weak Lensing, and Galaxy Correlations}

\author{C.~To}
\email[]{chto@stanford.edu}
\affiliation{Department of Physics, Stanford University, 382 Via Pueblo Mall, Stanford, CA 94305, USA}
\affiliation{Kavli Institute for Particle Astrophysics \& Cosmology, P. O. Box 2450, Stanford University, Stanford, CA 94305, USA}
\affiliation{SLAC National Accelerator Laboratory, Menlo Park, CA 94025, USA}
\author{E.~Krause}
\email[]{krausee@arizona.edu}
\affiliation{Department of Astronomy/Steward Observatory, University of Arizona, 933 North Cherry Avenue, Tucson, AZ 85721-0065, USA}
\affiliation{Department of Physics, University of Arizona, Tucson, AZ 85721, USA}
\author{E.~Rozo}
\affiliation{Department of Physics, University of Arizona, Tucson, AZ 85721, USA}
\author{H.~Wu}
\affiliation{Center for Cosmology and Astro-Particle Physics, The Ohio State University, Columbus, OH 43210, USA}
\affiliation{Department of Physics, Boise State University, Boise, ID 83725, USA}
\author{D.~Gruen}
\affiliation{Department of Physics, Stanford University, 382 Via Pueblo Mall, Stanford, CA 94305, USA}
\affiliation{Kavli Institute for Particle Astrophysics \& Cosmology, P. O. Box 2450, Stanford University, Stanford, CA 94305, USA}
\affiliation{SLAC National Accelerator Laboratory, Menlo Park, CA 94025, USA}
\author{R.~H.~Wechsler}
\affiliation{Department of Physics, Stanford University, 382 Via Pueblo Mall, Stanford, CA 94305, USA}
\affiliation{Kavli Institute for Particle Astrophysics \& Cosmology, P. O. Box 2450, Stanford University, Stanford, CA 94305, USA}
\affiliation{SLAC National Accelerator Laboratory, Menlo Park, CA 94025, USA}
\author{T.~F.~Eifler}
\affiliation{Department of Astronomy/Steward Observatory, University of Arizona, 933 North Cherry Avenue, Tucson, AZ 85721-0065, USA}
\author{E.~S.~Rykoff}
\affiliation{Kavli Institute for Particle Astrophysics \& Cosmology, P. O. Box 2450, Stanford University, Stanford, CA 94305, USA}
\affiliation{SLAC National Accelerator Laboratory, Menlo Park, CA 94025, USA}
\author{M.~Costanzi}
\affiliation{INAF-Osservatorio Astronomico di Trieste, via G. B. Tiepolo 11, I-34143 Trieste, Italy}
\affiliation{Institute for Fundamental Physics of the Universe, Via Beirut 2, 34014 Trieste, Italy}
\author{M.~R.~Becker}
\affiliation{Argonne National Laboratory, 9700 South Cass Avenue, Lemont, IL 60439, USA}
% ...

\collaboration{DES Collaboration}
 \date{\today}

\begin{abstract}
Combining multiple observational probes is a powerful technique to provide robust and precise constraints on cosmological parameters. In this \emph{letter}, we present the first joint analysis of cluster abundances and auto/cross correlations of three cosmic tracer fields measured from the first year data of the Dark Energy Survey: galaxy density, weak gravitational lensing shear, and cluster density split by optical richness. From a joint analysis of cluster abundances, three cluster cross-correlations, and auto correlations of galaxy density, we obtain $\Omega_{\rm{m}}=0.305^{+0.055}_{-0.038}$ and $\sigma_8=0.783^{+0.064}_{-0.054}$. This result is consistent with constraints from the DES-Y1 galaxy clustering and weak lensing two-point correlation functions
for the flat $\nu\Lambda$CDM model.
We thus combine cluster abundances and all two-point correlations from three cosmic tracer fields and find improved constraints on cosmological parameters as well as on the cluster observable--mass scaling relation. This analysis is an important advance in both optical cluster cosmology and multi-probe analyses of upcoming wide imaging surveys.
\keywords{Cosmology, Cosmological parameters, Galaxy cluster counts,  Large-scale structure of the universe}
\pacs{98.80.-k, 98.80.Es, 98.65.-r}
\end{abstract}
\maketitle 
\maketitle 
\noindent \textbf{\emph{Introduction.}}
--- The standard vacuum dark energy, cold dark matter (flat $\Lambda$CDM) cosmological model with just six parameters has been remarkably successful at describing a broad range of cosmological observations across the history of the universe. %
However, a fundamental physics explanation of the two main constituents of this model --- dark matter and dark energy --- is still missing.
This has inspired ambitious cosmic surveys that are testing the $\Lambda$CDM model with increasingly precise measurements of complementary cosmological probes \cite{Weinberg_2013}.

Wide-field imaging surveys, such as the Dark Energy Survey (DES\footnote{\href{https://www.darkenergysurvey.org/}{https://www.darkenergysurvey.org/}}), the Hyper-Suprime Cam Subaru Strategic Program (HSC\footnote{\href{http://www.naoj.org/Projects/HSC/HSCProject.html}{http://www.naoj.org/Projects/HSC/HSCProject.html}}), and the Kilo Degree Survey (KiDS\footnote{\href{http://www.astro-wise.org/projects/KIDS/}{http://www.astro-wise.org/projects/KIDS/}}), are one class of these cosmic surveys, which map the spatial distribution, shapes, and colors of millions of galaxies. These wide-field imaging data sets enable a wide range of cosmological measurements \cite{DESmethod, DESY1KP, HSC1,HSC2,KIDScombine,KiDSCombine2,heymans2020kids1000}. Two of the most established cosmological probes are galaxy clustering and weak gravitational lensing. Analyses that include the auto-correlation of these two tracer fields as well as their cross correlation, galaxy--galaxy lensing, are referred to as \ttt\ analyses and are emerging as a competitive cosmological test.

The abundances and spatial distribution of galaxy clusters, which are associated with the highest peaks in the matter density field, provides another powerful probe of cosmic structure formation and expansion history \cite{Allen11}. Clusters can be detected in wide-field imaging data as associations of large numbers of galaxies. Confronting observations of galaxy clusters with predictions of the $\Lambda$CDM model requires an understanding of the observational selection of clusters and the relation between observed cluster properties and the total cluster mass. The latter is characterized as the mass--observable relation (MOR), and presents one of the key modeling challenges for unlocking the potential of cluster cosmology \cite{WtG1,WtG4, SPTClusters2019, matteoSDSS, DES_cluster_cosmology}. In this work, we combine three cluster related \cczs\ with galaxy clustering to calibrate the MOR. The combination of these four two-point correlation functions is expected to yield a precise measurement of cluster biases relative to matter density fluctuations  
\cite{2020MNRAS.491.3061S, Simpaper}, from which a reliable cluster mass--observable relation can be obtained \cite{halomodel,Tinker10, Baxter16,2020arXiv200513564C}. Thus, the combination of these four two-point correlations and cluster abundances, referred to as a \clustercomb\ analysis, can yield competitive cosmological constraints \citep{Simpaper}. We note that most of the cosmological information in the \clustercomb\ analysis comes from cluster abundances, while the additional two point functions combine to break degeneracies with the mass--observable relation; therefore, we consider it as a cluster cosmology analysis.

\begin{figure}[ht!]
\centering
\includegraphics[width=0.5\textwidth]{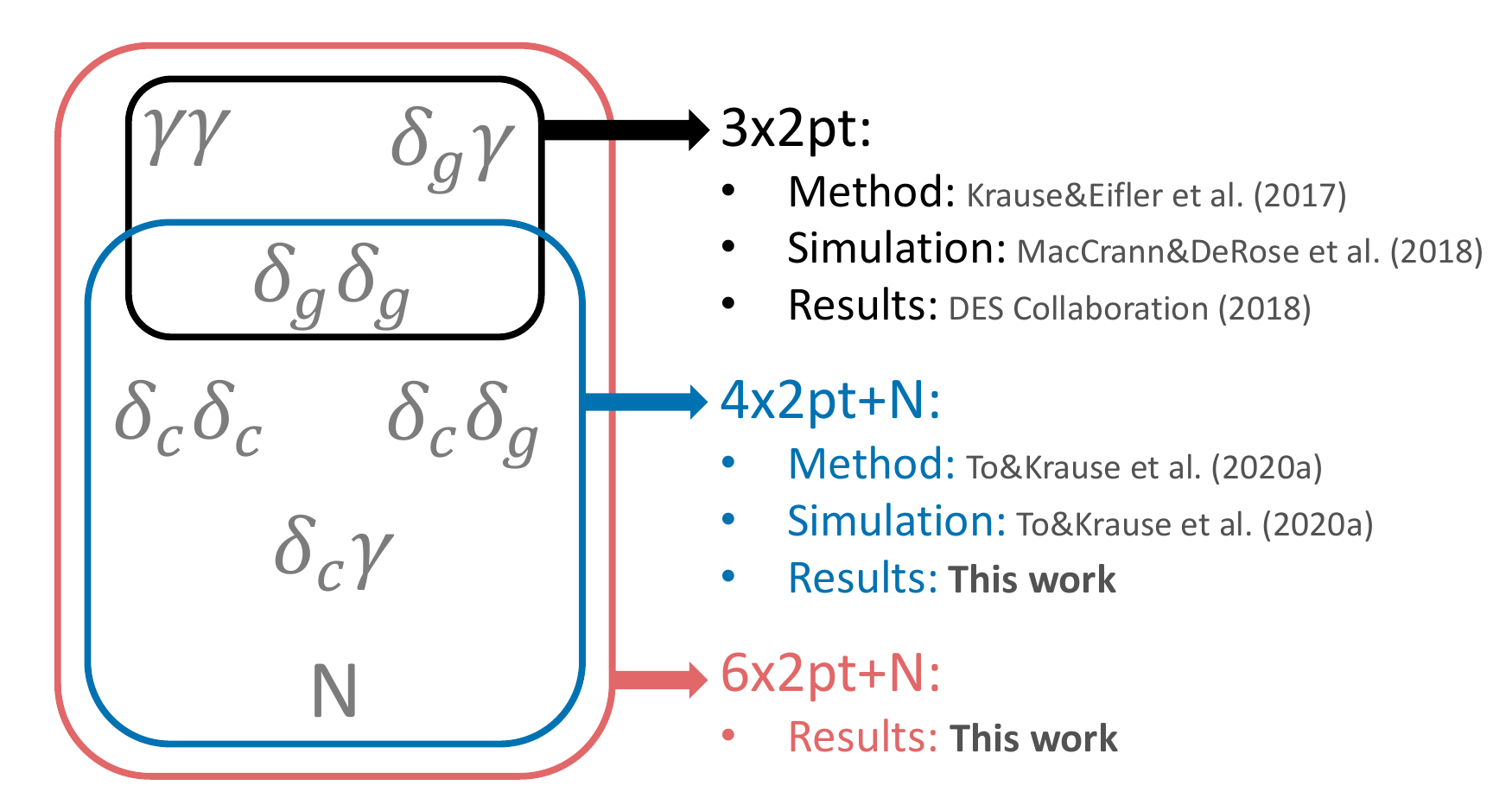}\hspace{-0.05\textwidth}
\caption{Summary of the different components in this analysis and a non-exhaustive list of papers describing and validating the method adopted in this analysis. For a more comprehensive list of papers this work relies on, we refer the reader to \citep[and references therein]{DESY1KP, DES_cluster_cosmology}. The data in this paper consist of cluster abundances (N) and six two-point correlation functions derived from three cosmic tracer fields, namely galaxy density ($\delta_g$), weak gravitational lensing shear ($\gamma$), and cluster density ($\delta_c$). The correlation functions include cosmic shear ($\gamma \gamma$), galaxy--galaxy lensing ($\delta_g \gamma$), galaxy clustering ($\delta_g \delta_g$), cluster--galaxy cross-correlation ($\delta_c \delta_g$), cluster auto-correlation ($\delta_c \delta_c$), and cluster lensing ($\delta_c \gamma$). The black box denotes the joint \ttt\ analysis. The blue and red boxes represent the \clustercomb\ and \allcomb\ analyses presented in this \emph{letter}.}
\label{fig:summarydata}
\end{figure}

In this \emph{letter}, we first demonstrate the consistency between our cluster cosmology analysis (\clustercomb), the \ttt\ analysis, and other cluster cosmology analyses, in the context of the $\Lambda$CDM model with massive neutrinos ($\nu\Lambda$CDM). We then present the first \emph{joint} analysis, referred to as \allcomb, of galaxy clusters abundances and clustering, galaxy clustering, and weak gravitational lensing. 
In Fig.~\ref{fig:summarydata}, we summarize the different components of the analysis.  
Our analysis uses %
the same set of systematics modeling, calibration procedures, and analysis pipeline across all probes, and properly accounts for the covariance between the probes.
We demonstrate that combining galaxy clusters and the \ttt\ analysis improves both cosmological and cluster mass--observable relation constraints, compared to these individual analyses. 

\noindent \textbf{\emph{Data and Measurement.}}
---
We measure galaxy density fields, weak gravitational lensing shear fields, and  cluster density fields from the $1321$ $\rm{deg}^2$ of imaging data taken in the first season of the Dark Energy Survey \cite{DES} (DESY1). The measurement is based on procedures described in \cite{Simpaper} using the DESY1 public catalogs\footnote{\href{https://des.ncsa.illinois.edu/releases/y1a1/key-catalogs}{https://des.ncsa.illinois.edu/releases/y1a1/key-catalogs}}. These include the \redmagic\ galaxy catalog \citep{Redmagic} for the galaxy density field; the \textsc{METACALIBRATION} shape catalog \citep{2018MNRAS.481.1149z} and \textsc{BPZ} photometric redshift (photo-$z$) catalog \citep{photoz} for the weak gravitational lensing shear field; and the \redmapper\ cluster catalog \citep{Redmapper1} for the cluster density field. To construct the galaxy density field, $\sim 650,000$ \redmagic{} galaxies over the redshift range $0.15<z<0.9$ are split into five redshift bins based on their photo-$z$ estimations. The weak gravitational lensing shear field is constructed based on $\sim26$ million galaxies spanning the redshift range $0.2<z<1.3$, split into four redshift bins based on \textsc{BPZ} photo-$z$ estimation. For the cluster density fields, $4794$ \redmapper\ clusters are split into three redshift bins spanning from redshift range $0.2<z<0.6$. In each redshift bin, these clusters are further split into four bins based on their richness ($\lambda$), a cluster mass proxy defined as a weighted sum of the cluster red-sequence member galaxies. These clusters span the richness range $20<\lambda<235$.

Six two-point correlations are constructed from the three cosmic tracer fields, namely the shear auto-correlation (cosmic shear), the galaxy position--shear \ccz\ (galaxy--galaxy lensing), the galaxy position auto-correlation (galaxy clustering), the galaxy position--cluster position cross correlation (galaxy--cluster cross-correlation), the cluster position-shear \ccz\ (cluster lensing), and the cluster position auto-correlation (cluster clustering). The first three two-point correlation functions are the DESY1 public \ttt\ data vector\footnote{\href{https://des.ncsa.illinois.edu/releases/y1a1/key-products}{https://des.ncsa.illinois.edu/releases/y1a1/key-products}}. The last three two-point correlations and cluster abundances are measured following procedures described in \cite{Simpaper}. 

\noindent\textbf{\emph{Modeling + Inference.}}
---
To analyze the observed data vectors, we assume a Gaussian likelihood function which requires a covariance matrix and a theory model. The details of constructing these two components are specified below.  
\\
\emph{Covariance and Model}
---
The covariance matrix \cite{supp} is derived based on halo models \citep{halomodel,cosmolike2016} and is validated in \cite{DESmethod, Simpaper}. The derivation and construction procedures are detailed in \cite{Simpaper}. In terms of the theory modeling, we relate the abundances of galaxy clusters to the halo mass function \citep{Tinker10} assuming a power-law relation with log-normal scatter between the halo mass and cluster richness \citep{Simpaper}.
The three cosmic tracer fields are assumed to be linearly connected to the matter density fields, which are modeled using \textsc{CLASS} \citep{CLASS} and \textsc{Halofit} \citep{Takahashi2012}. The model of cosmic shear and galaxy--galaxy lensing is described and validated in \cite{DESmethod, Niall}, while the model of \clustercomb\ is described and validated in \cite{Simpaper} with modifications to the modeling of the effect of massive neutrinos \cite{supp}. Both the covariance matrix derivation and the model prediction are implemented in \textsc{CosmoLike} \citep{cosmolike2016}\footnote{The version used in this work is tagged as 'cluster$\_$chto' in the 'cosmolike$\_$core' github repository and 'desy1$\_$paper' in the 'lighthouse' repository of the \textsc{CosmoLike} github organization.}.
\\
\emph{Analysis Choices} 
---
In addition to the model and covariance details described above, we have designed our analysis to ensure robustness of the inferred result. We summarize the key analysis choices below. 
\begin{enumerate}[label=(\emph{{\roman*}})]
\item \emph{Only large scale information is used.} Due to uncertainties of modeling baryonic effects, non-linear relations between cosmic tracer fields and matter density fields, and random fluctuations of sparse tracers on small scales, 
we adopt conservative angular scale cuts on the two-point correlation functions. The scale cuts of \ttt\ data vectors are defined and validated in \cite{DESmethod}; the scale cuts of \clustercomb\ are defined and validated in \cite{Simpaper}. 
\item \emph{The same set of parameters and priors are used in \ttt, \clustercomb, and \allcomb\ analyses.} In addition to the six cosmological parameters in the $\nu\Lambda$CDM model, we simultaneously sample over 26 additional nuisance parameters \cite{supp}. These include galaxy bias parameters (5), lens and source galaxy photo-$z$ biases (9), multiplicative shear biases (4), intrinsic alignment parameters (2), parameters describing the richness--mass relation (4), and parameters describing selection bias for clusters (2). For detailed descriptions of these nuisance parameters and the associated priors, we refer the readers to \cite{DESmethod, Simpaper, supp}. We note that we do not account for intrinsic alignments in the cluster lensing analysis. The effect is expected to be small \citep{2015A&A...575A..48S} and was not included in the previous weak lensing analysis of the same sample \citep{Tomclusterlensing}. In addition, in the cluster lensing model, we exclude bins where the maximum redshift of galaxy clusters is larger than the mean redshift of source galaxies.  
\item \emph{The analysis was done blindly.} Cosmological parameters were blinded by random shifts before the analysis choices were determined. We detail our blinding procedure in the Supplemental Material \cite{supp}.
\end{enumerate}

\begin{table}[t]
\begin{tabular}{lcccl}
\hline

Parameter & \ttt\ & \clustercomb\ &  \allcomb\ & Flat Prior \\\hline \hline\\[-0.5em]
$\omegam$ & $0.297\pm 0.036$  & $0.305^{+0.055}_{-0.038}  $ &$0.276^{+0.033}_{-0.026}     $ & [0.1,0.9]\\
$A_s$ ($\times 10^{9})$ & $2.15^{+0.38}_{-0.34}$ & $2.27^{+0.57}_{-0.41}$& $2.08^{+0.41}_{-0.31}$& [$0.5$, $5$] \\
$n_s$ &-&- &- & [0.87, 1.07]\\
$\Omega_{\rm{b}}$ & - & -& -& [0.03,0.07]\\
$\Sigma m_\nu [eV]$& - & -&- & [$0.047$, $0.931$]\\
$h$& -&- & -& [0.55, 0.91]\\
$\sigma_8$ & $0.771^{+0.064}_{-0.054}$&$0.783^{+0.064}_{-0.054}$ & $0.802^{+0.056}_{-0.048}$& Derived\\
\hline \hline\\[-0.5em]
$\chi^2$ (d.o.f) & 512 (444)& 610 (567) & 1054 (992)&\\
$p$-value  &0.014 & 0.103 & 0.084 & 
\\ \hline
\end{tabular}
\caption{
Summary of cosmological parameter constraints in the $\nu\Lambda CDM$ model from three combinations of data vectors: \ttt\ is a joint analysis of cosmic shear, galaxy--galaxy lensing, and galaxy clustering; \clustercomb\ stands for a joint analysis of cluster abundances, cluster--galaxy cross correlations, cluster clustering, and cluster lensing; \allcomb\ corresponds to a joint analysis of \ttt\ and \clustercomb.  The number reported is the 1D peak of the posterior and the asymmetric $68\%$ confidence interval. Cells with no entries correspond to posteriors dominated by the priors. The last two rows summarize the goodness of fit for each data vector computed at the best-fit model. 
}
\label{tab:summ_all}
\end{table}
\begin{figure}[t]
\centering
\includegraphics[width=0.5\textwidth]{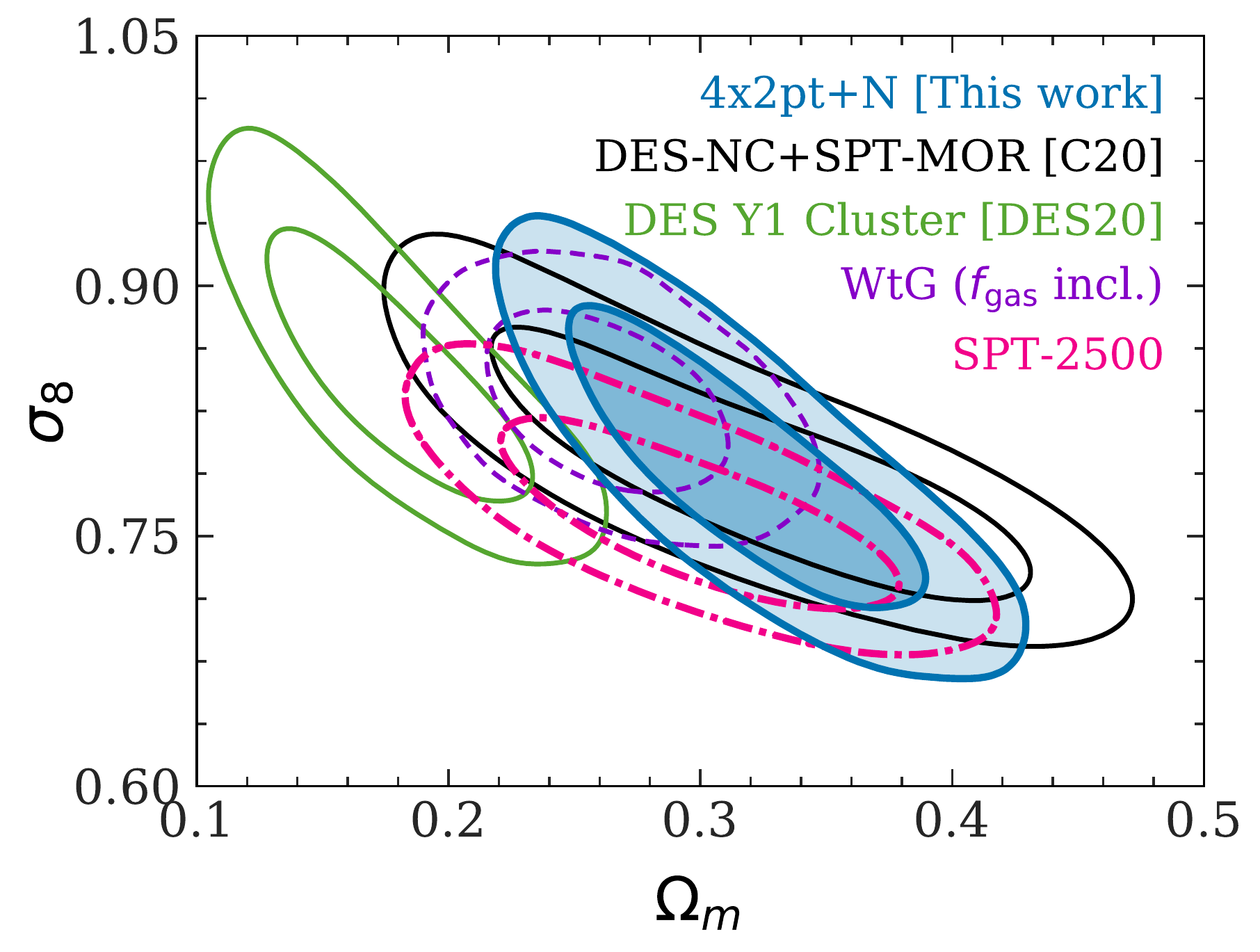}
\caption{Comparison of $\nu\Lambda$CDM constraints on $\omegam$ and $\sigmae$ derived from \clustercomb\ (blue) and other cluster cosmology analyses in the literature: DES-Y1 joint analysis of cluster abundances and weak lensing mass estimates from \cite{DES_cluster_cosmology} (green); a joint analysis of DES cluster abundances and SPT-SZ multi-wavelength data from \cite{Matteoawesomepaper} (black); the Weighing the Giants study from \cite{WtG4} (purple); the SPT-2500 analysis from \cite{SPTClusters2019} (pink). Contours show $68\%$ and $95\%$ confidence levels.}
\label{fig:compcluster}
\end{figure}
\noindent\textbf{\emph{Results and Discussions}}
---
Table~\ref{tab:summ_all} presents the cosmological parameter constraints from \ttt, \clustercomb, and \allcomb. 

\emph{Cluster cosmology}
--- 
We first compare our cosmological constraints (\clustercomb) with cluster analyses in the literature. The result is shown in Fig.~\ref{fig:compcluster}. According to the $Q_{\rm{DM}}$ tension metric \cite{Tension1},  the \clustercomb\ constraints agree with most of the cluster cosmology analyses within $0.6\sigma$, except for the constraints from a joint analysis of cluster abundances and weak lensing mass estimates in the DES-Y1 data \cite{DES_cluster_cosmology} (hereafter called DES20). The DES20 analysis is in $2.9 \sigma$ tension with our \clustercomb\ analysis despite the fact that the two analyses share the same galaxy cluster and weak gravitational lensing shear catalogs. The main difference between \clustercomb\ and DES20 is that \clustercomb\ only uses large-scale information while the DES20 signal-to-noise is dominated by small-scale cluster lensing. We note that a similar tension has been found when comparing DES20 with a joint analysis of the DES cluster abundances and SPT-SZ multi-wavelength data \cite{Matteoawesomepaper} (hereafter called C20). In C20, the cluster mass--observable scaling relation is calibrated by cross-matching the \redmapper{} and SPT-SZ catalog (mean $\lambda=78$) and using the high-quality X-ray and weak lensing follow-up data available for 121 SPT-SZ clusters to constrain the scaling relation \citep{Bleem15,Schrabback18,Dietrich2019, McDonald2013, McDonald2017,Bocquet2019}. %
Comparison between DES20, C20, and \clustercomb\ suggests that the tension between the DES20 analysis and other cluster cosmology analyses is likely due to unmodeled systematic artifacts in the weak lensing data of the \redmapper\ clusters at small scales, as it is precisely this component of the data that we ignore. This is consistent with the interpretation advanced by DES20.  The low lensing signal observed for \redmapper\ clusters may be related to the lensing-is-low problem for massive galaxies in the SDSS \cite{2017MNRAS.467.3024L}: the prediction of the best-fit model from galaxy clustering is larger than the measured galaxy--galaxy lensing signal. Should these two lensing anomalies be related, it is interesting to note that this anomaly seems to disappear at the high mass end of the mass function.  The correct resolution to this lensing anomaly at small scales remains to be seen.

\begin{figure}[t]
\raggedright
\includegraphics[width=0.5\textwidth]{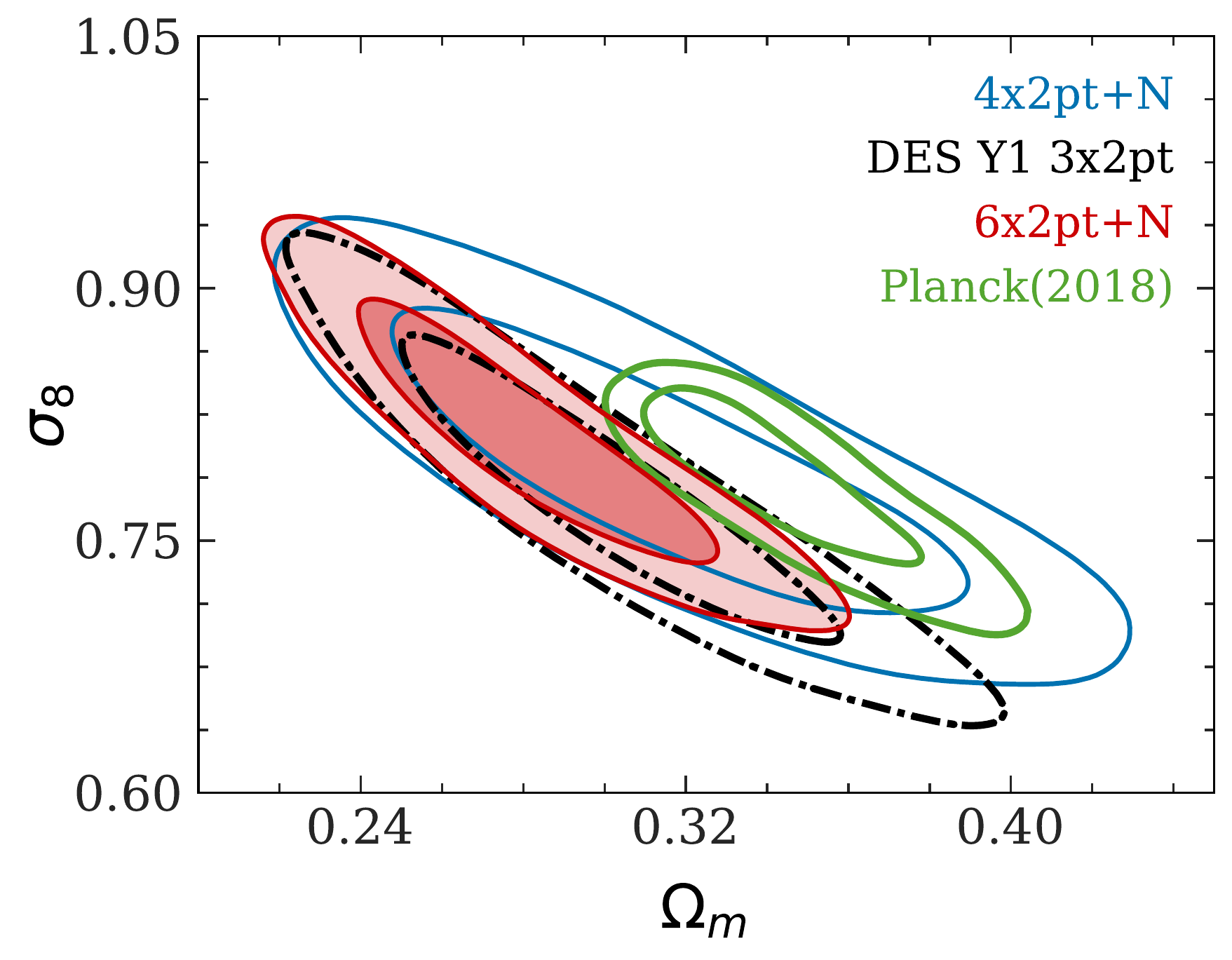}
\caption{$\nu\Lambda$CDM constraints on $\omegam$ and $\sigmae$ from \ttt\ (black), \clustercomb\ (blue), and their combination (red). For comparison, the green contours show constraints from the CMB at high redshift (Planck without lensing). Contours show $68\%$ and $95\%$ confidence levels.}
\label{fig:alldes}
\end{figure}
\begin{figure*}[t]
\raggedright
\includegraphics[width=1.0\textwidth]{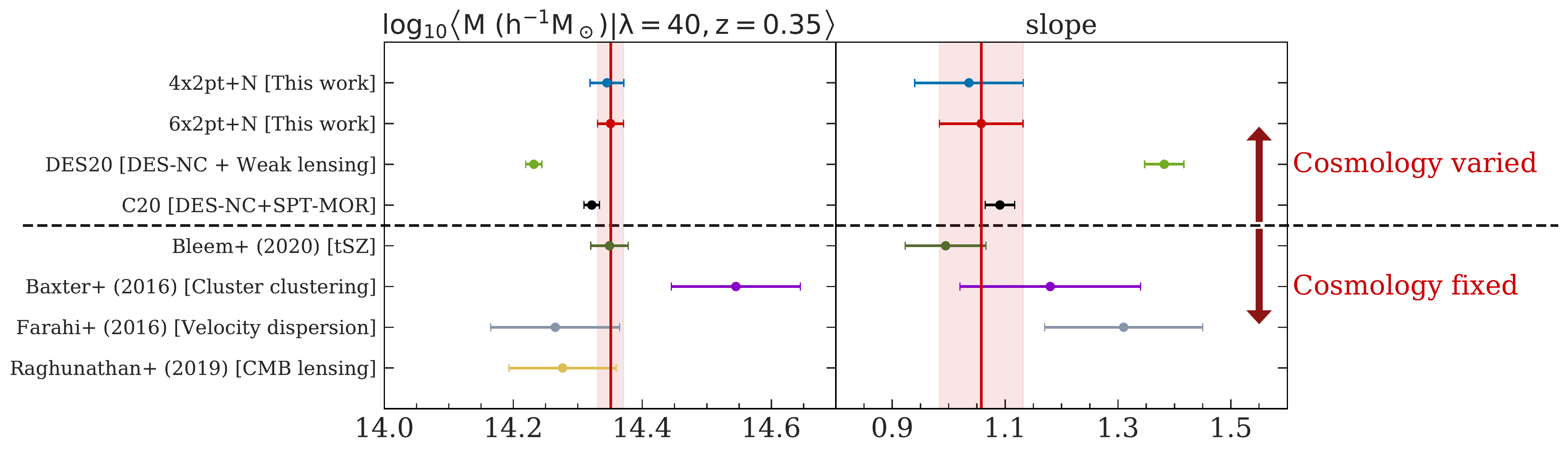}\hspace{-0.05\textwidth}
\caption{Comparison of the predicted mean mass at richness $\lambda=40$ and redshift $z=0.35$ and the slope of the richness scaling relation from this \emph{letter} (blue and red) with results in literature: a joint analysis of number counts and weak lensing mass estimates \cite{DES_cluster_cosmology} (light green); %
a joint analysis of DES cluster abundances and SPT-SZ multi-wavelength data \cite{Matteoawesomepaper} (black);
SZ scaling relation \cite{Bleem19} (dark green); auto-correlations of galaxy clusters \cite{Baxter16} (purple); velocity dispersion \cite{Farahi16} (gray); and CMB lensing \cite{Raghunathan19} (brown). Error bars show $68\%$ confidence intervals. The slope is unconstrained by CMB lensing.}
\label{fig:meanmass}
\end{figure*}

\emph{Systematics of \redmapper\ clusters}
--- 
Photometrically selected galaxy clusters are subject to two important systematics: projection effects \citep{Matteoprojection, Tomomi, DES_cluster_cosmology} and orientation biases \citep{Heidiorientation, DES_cluster_cosmology}. These two systematics bias the observed galaxy and matter overdensities of the selected galaxy clusters relative to randomly selected halos of the same mass. On large scales these two effects manifest as an additional bias factor ($b_{\rm{sel}}$) in the amplitude of the correlation functions, which can be sufficiently described by a power law in mass: $b_{\rm{sel}}(M)=b_{s0}(M/5\times10^{14} h^{-1}M_\odot)^{b_{s1}}$\cite{Simpaper}. From the \allcomb\ analysis, we obtain $b_{s0}=1.15^{+0.11}_{-0.09}$ and $b_{s1}=-0.029^{+0.056}_{-0.062}$. Comparing this constraint with predictions from simulations and theory might shed light on important systematics of photometrically selected galaxy clusters. We leave these interesting comparisons to future studies.

\emph{Comparison of different cosmological probes in the Dark Energy Survey}
---
Fig.~\ref{fig:alldes} shows a comparison between \ttt\  and \clustercomb. Here, before the analysis was unblinded, the tension metric was set to $Q_{\rm{UDM}}$ \cite{Tension1, Tension2}, which compares the parameters from \ttt\ and from its combination with \clustercomb.  According to $Q_{\rm{UDM}}$, the tension between  \ttt\  and \clustercomb\ is $0.024\sigma$, indicating a strong consistency between galaxy clustering, weak gravitational lensing, and galaxy clusters in the context of the $\nu\Lambda CDM$ model. Given the demonstrated consistency between \ttt\ and \clustercomb, we proceed to perform a joint analysis of cluster abundances and all six two-point correlation functions derived from galaxy density fields, galaxy cluster density fields, and weak gravitational lensing shear fields. The constraints from this combination (\allcomb) are shown in Fig.~\ref{fig:alldes}. Evidently, our \allcomb\ analysis leads to a $\sim 20\%$ improvement on the constraints of $\omegam$, the energy density of total matter in the universe, compared to the constraints from \ttt. Since DES only measures the matter distribution when the universe is older than $10$ billion years, it is interesting to compare our constraints on $\omegam$ and $\sigma_8$ with constraints from the early universe --- this provides a consistency test of the $\nu\Lambda$CDM model across cosmic epochs. Specifically, we compare our result with the prediction from the joint TT, EE, BB, TE likelihood measured by the Plank satellite \citep{Planck2018}, reanalyzed using the DES analysis choice of marginalizing over the unknown sum of neutrino masses \citep{DESY1KP}. The comparison is shown in Fig.~\ref{fig:alldes}. Despite the visual offset between Planck $\nu\Lambda$CDM prediction and \allcomb, we find that the tension is at the level of $1.42\sigma$ according to the tension metric \cite{Tension3}, which was set before the analysis was unblinded. The consistency between \allcomb\ and Planck is strong confirmation of the validity of the $\nu\Lambda$CDM model. Built on many previous works \cite[and references therein]{DESY1KP, DES_cluster_cosmology},  Fig.~\ref{fig:alldes} presents the first joint analysis of galaxy clustering, galaxy lensing, and galaxy clusters and is an important milestone in multi-probe analyses of wide-field imaging surveys.

\emph{Mean mass of \redmapper{} clusters}
---
A precise measurement of cluster masses is important, for cosmological exploitation of cluster samples as well as for astrophysical studies involving galaxy clusters \citep[e.g.][]{2019SSRv..215...25P, 2019MNRAS.487.2900S,2020ApJ...897...15T,2020MNRAS.494.1705G}. From \clustercomb\ and \allcomb\ analyses, we can derive the mean mass of the \redmapper{} clusters and its dependence on the richness. The result is shown in Fig.~\ref{fig:meanmass} and the calculation is detailed in \citep{supp}. The combination of clusters and \ttt\ yields a $\sim 20\%$ improvement on the constraints of mean cluster masses and their richness dependency compared to \clustercomb.  From the \allcomb\ analysis, the mean mass of \redmapper\ clusters at $z=0.35$ is constrained as %
\begin{equation}
    \langle M_{200\rm{m}} | \lambda \rangle = 10^{14.351 \pm 0.020}\left(\frac{\lambda}{40}\right)^{1.058\pm 0.074} \nonumber h^{-1}M_{\odot}, 
\end{equation}
where $M_{200\rm{m}}$ is the mass enclosed within a sphere in which the mean matter density is equal to 200 times the mean matter density of the universe.
In Fig.~\ref{fig:meanmass}, we compare our constraints with results in the literature and find that our constraints are competitive with these results, while properly marginalizing over cosmological parameters. The result herein is consistent with C20 despite many differences between the two analyses. These differences include scale cuts: \clustercomb\ only uses scales greater than $8h^{-1}\rm{Mpc}$, while C20 only uses small-scale lensing for mass calibration; mass ranges: \clustercomb\ uses all \redmapper{} clusters with $\lambda>20$, while C20 only uses high richness system (mean $\lambda=78$) from the \redmapper{}-SPT-SZ cross-matched sample for the mass calibration; differences in the data: \clustercomb\ only uses data from the optical surveys for mass calibrations, while C20 uses the high-quality X-ray and weak lensing follow-up data available for 121 SPT-SZ clusters for mass calibrations. The consistency between the two analyses demonstrates the robustness of the mass constraints. We note that constraints on the mean mass and the slope of the mass--richness relation can be sensitive to assumptions about the projection modeling \cite{Matteoawesomepaper}; this will be an interesting direction for future investigation. 

\noindent\textbf{\emph{Conclusions and outlook}}
---
Combining multiple cosmological probes has long been advocated as a promising avenue to constrain cosmological parameters. Different probes are sensitive to different aspects of cosmic structure formation and are affected by different astrophysical uncertainties.  However, combining different cosmological probes from the same survey faces many challenges. First, probes that involve different tracers of the large-scale structure are correlated, since they probe the same dark matter density field. Second, different probes are affected by the same systematic errors, such as intrinsic alignments and photometric redshift uncertainties. Thus, a joint analysis of different cosmological probes requires a consistent modeling of systematics and statistical uncertainties to accurately capture the cosmological information content of wide-field imaging surveys. 

In this \emph{letter}, we present the first joint analysis of cluster abundances and six two-point correlation functions derived from three cosmic tracer fields: galaxy density, weak gravitational lensing shear, and cluster density. 
Our findings can be summarized as follows:
\begin{enumerate}[label=(\emph{{\roman*}})]
\item Despite the surprising results of the DES-Y1 cluster abundances analysis \citep{DES_cluster_cosmology}, our multi-probe cluster cosmology approach based on photometrically selected samples yields cosmological constraints that are consistent with other cluster cosmology analyses and other cosmological probes in DES. This is likely a consequence of our analysis being restricted to large scales only. This result, together with C20 \cite{Matteoawesomepaper}, suggests that the modeling of small-scale cluster lensing for low mass optically selected clusters is currently insufficient and is likely a cause of the biased cosmology in \citep{DES_cluster_cosmology}. 

\item We find that combining galaxy clusters  with galaxy clustering and weak gravitational lensing improves both cosmological constraints and constraints on the mean mass of galaxy clusters by $\sim 20\%$, compared to results from analyses of individual probes.

\item The combined cosmological constraint from DES is consistent with Planck at the $1.4\sigma$ level in the context of the $\nu\Lambda$CDM model. 

\item Combining galaxy clusters with galaxy clustering and weak gravitational lensing provides a precise constraint on the mean mass of galaxy clusters and its richness dependence. 

In the near future, we expect a $\sim 40\%$ improvement in cosmological constraints for \clustercomb\ from the analysis of the first three years of data from the Dark Energy Survey, mostly due to the increased survey area. This improvement will be followed by significant additional improvements from upcoming wide imaging surveys in the 2020s \cite{LSST, Euclid, Wfirst}. The analysis presented in this \emph{letter} is an important step towards fully realizing the potential of these richer and larger datasets. 

\end{enumerate}
\vspace{1em} \noindent \textbf{\emph{Acknowledgements.}}
---
This paper has gone through internal review by the DES collaboration. This work was supported in part by the U.S. Department of Energy contract to SLAC National Accelerator Laboratory, under contract no. DE-AC02-76SF00515 (CH, DG, RW) including a Panofsky Fellowship awarded to DG. EK is supported by the Department of Energy grant DE-SC0020247. ER is supported by DOE grants DE-SC0015975 and DE-SC0009913, and by NSF grant 2009401.  ER also acknowledges funding from the Cottrell Scholar program of the Research Corporation for Science Advancement. HW is supported by NSF Grant AST-1516997. Some of the computing for this project was performed on the Sherlock cluster. We would like to thank KIPAC, Stanford University, and the Stanford Research Computing Center for providing computational resources and support that contributed to these research results.

Funding for the DES Projects has been provided by the DOE and NSF(USA), MEC/MICINN/MINECO(Spain), STFC(UK), HEFCE(UK). NCSA(UIUC), KICP(U. Chicago), CCAPP(Ohio State), 
MIFPA(Texas A\&M), CNPQ, FAPERJ, FINEP (Brazil), DFG(Germany) and the Collaborating Institutions in the Dark Energy Survey.

The Collaborating Institutions are Argonne Lab, UC Santa Cruz, University of Cambridge, CIEMAT-Madrid, University of Chicago, University College London, 
DES-Brazil Consortium, University of Edinburgh, ETH Z{\"u}rich, Fermilab, University of Illinois, ICE (IEEC-CSIC), IFAE Barcelona, Lawrence Berkeley Lab, 
LMU M{\"u}nchen and the associated Excellence Cluster Universe, University of Michigan, NFS's NOIRLab, University of Nottingham, Ohio State University, University of 
Pennsylvania, University of Portsmouth, SLAC National Lab, Stanford University, University of Sussex, Texas A\&M University, and the OzDES Membership Consortium.

Based in part on observations at Cerro Tololo Inter-American Observatory at NSF’s NOIRLab (NOIRLab Prop. ID 2012B-0001; PI: J. Frieman), which is managed by the Association of Universities for Research in Astronomy (AURA) under a cooperative agreement with the National Science Foundation.

The DES Data Management System is supported by the NSF under Grant Numbers AST-1138766 and AST-1536171. 
The DES participants from Spanish institutions are partially supported by MICINN under grants ESP2017-89838, PGC2018-094773, PGC2018-102021, SEV-2016-0588, SEV-2016-0597, and MDM-2015-0509, some of which include ERDF funds from the European Union. IFAE is partially funded by the CERCA program of the Generalitat de Catalunya.
Research leading to these results has received funding from the European Research
Council under the European Union's Seventh Framework Program (FP7/2007-2013) including ERC grant agreements 240672, 291329, and 306478.
We  acknowledge support from the Brazilian Instituto Nacional de Ci\^encia
e Tecnologia (INCT) do e-Universo (CNPq grant 465376/2014-2).

This manuscript has been authored by Fermi Research Alliance, LLC under Contract No. DE-AC02-07CH11359 with the U.S. Department of Energy, Office of Science, Office of High Energy Physics. 
\bibliographystyle{apsrev}
\bibliography{sample}
\clearpage
\onecolumngrid
\appendix

\onecolumngrid
\appendix

\section{SUPPLEMENTAL MATERIAL}
\subsection{Blinding Strategy}
The blinding strategy aims to avoid confirmation biases resulting from adjusting the analysis strategy based on one's expectation of the outcome. In this paper, we cannot claim a fully blinded analysis, because \redmagic{} clustering and \redmapper{} number counts were unblinded in the \cite{DESY1KP} and \cite{DES_cluster_cosmology} analyses. However, given that we add new components to the data vectors and make different analysis choices from \cite{DES_cluster_cosmology}, we decide to perform a blinded cosmology analysis, in which the following protocols are followed:
\begin{enumerate}[label=(\emph{{\roman*}})]
    \item The cosmological parameters and richness--mass relation parameters in MCMC were randomly displaced before being stored. 
    \item All of the analysis choices are set by analyzing mock Dark Energy Surveys \cite{Simpaper}, except for the modeling of massive neutrinos. The modeling of massive neutrinos is described in the Supplement Material, and this analysis choice is decided before unblinding. 
    \item All priors are set before unblinding. 
    \item We commit to providing the results obtained immediately after unblinding. 
    \item We commit to describing all post-unblinding analyses in a separate section. Note that after unblinding, we found no need for post-unblinding analyses.
\end{enumerate}    
We decide to unblind our analysis when the following criteria are fulfilled:
\begin{enumerate}[label=(\emph{{\roman*}})]
\item Our inference pipeline has to successfully recover the input cosmology in synthetic data sets. 
\item The analysis using mock DES data \cite{Simpaper} is approved by DES internal reviewers and the DES Cluster/Theory and Combined probe/Weak Lensing/Simulation working groups. 
\item We present the measurements, analysis choices, simulated analyses, and blinding strategy to three DES internal reviewers, who must explicitly agree that the analysis is ready to unblind.
\end{enumerate}
\subsection{Modeling of massive neutrinos}
In \cite{Simpaper}, we show that the Tinker halo mass function and Tinker bias model \citep{Tinker10} are sufficient for the analysis of the 4x2pt+N data vector. However, these models are established on N-body simulations without massive neutrinos; thus, they might not be suitable for an analysis of real data.  Several studies have investigated how the halo mass function and halo bias model can be extended to incorporate the effect of massive neutrinos \citep[e.g.][]{neutrino3, neutrino1,neutrino2}. While massive neutrinos suppress structures on scales smaller than their free-streaming length, they contribute negligibly to the collapse of massive dark matter halos. Thus, the halo mass function and halo bias are expected to depend only on cold dark matter (CDM) and baryons. In fact, \cite{neutrino3} and \cite{neutrino1} find that by including only CDM and baryons, the Tinker halo mass function and Tinker bias model provide a good description to measurements in simulations with massive neutrinos. Following these works, we modify our halo mass function and halo bias model as follows. 

The halo mass function is defined as 
\begin{equation}
\frac{dn}{dM} = f^{\rm{Tinker}}(\nu) \frac{\rho_{\rm{cdm+b}}}{M}\frac{d\rm{ln}\nu}{dM},
\end{equation}
where
\begin{align}
\label{eq:nu}
\nu &= \frac{\delta_c}{\sigma(R)},\\
\sigma^2(R) &= \int P_{\rm{cdm+b}, \rm{lin}}(k)W^2(kR)k^2dk,
\end{align}
$f^{\rm{Tinker}}(\nu)$ is the Tinker fitting formula \citep{Tinker10}, and $P_{\rm{cdm+b}, \rm{lin}}(k)$ is the linear CDM + baryon power spectrum as a function of wavenumber $k$. $W(kR)$ is the Fourier transform of the real space top-hat window function of radius $R$, which is defined as 
\begin{equation}
    R = \left(\frac{3M}{4\pi\rho_{\rm{cdm+b}}}\right) ^{\frac{1}{3}}.
\end{equation}

In this analysis, the bias model is the Tinker bias \citep{Tinker10}  using $\nu$ defined in equation~\ref{eq:nu}. Although \cite{neutrino1,neutrino2} find that relating halos to the total matter distribution in simulations with massive neutrinos results in a scale-dependent bias on large scales, \cite{DESmethod} find that such scale-dependent bias has negligible impact on cosmological constraints at DES Y1 accuracy. To be consistent with \cite{DESY1KP}, we relate two-point correlations to the total matter power spectrum in the universe.

\subsection{Blind tests}
\begin{figure}
\centering
\includegraphics[width=0.5\textwidth]{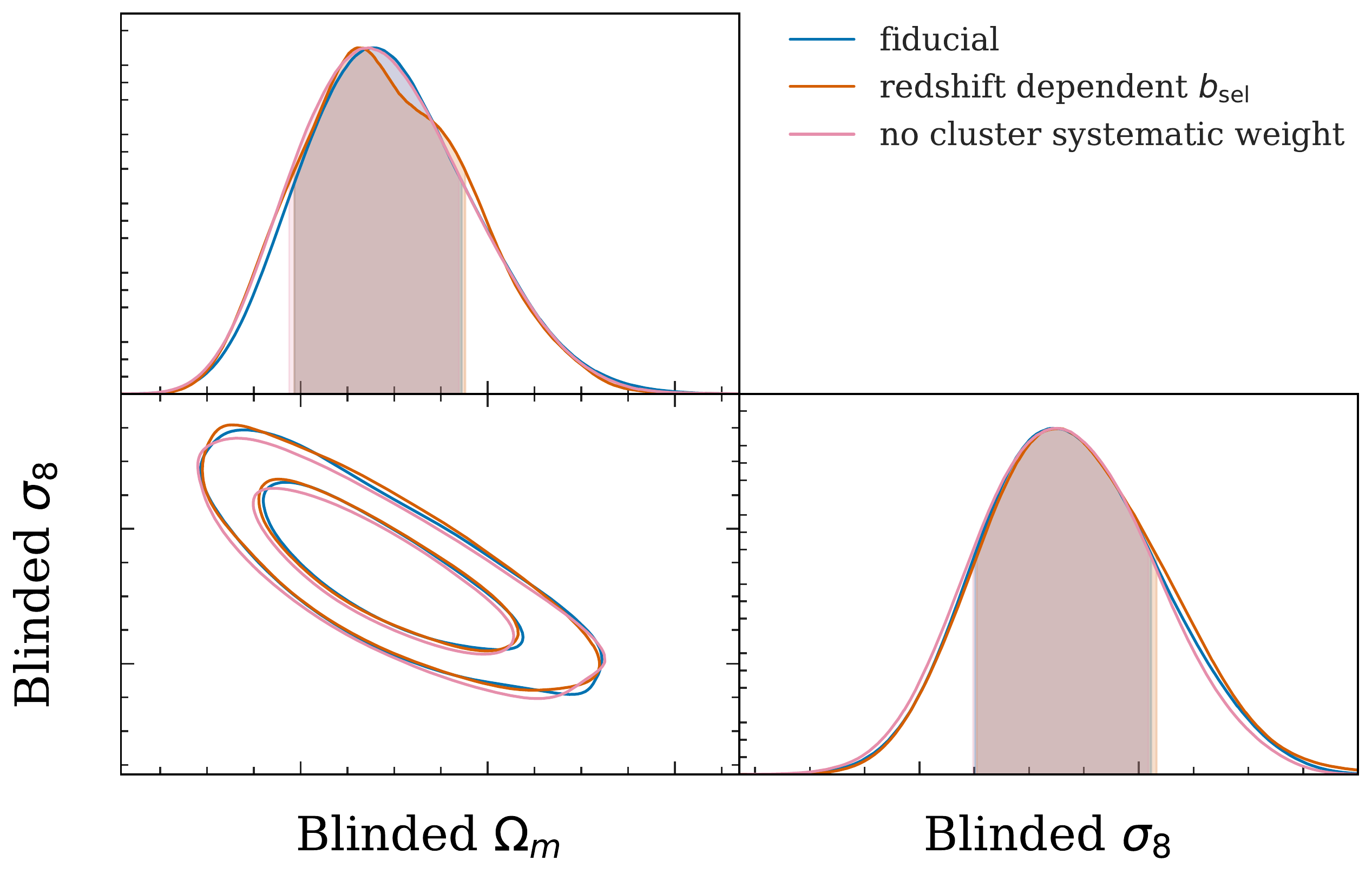}\hspace{-0.05\textwidth}
\caption{Blinded constraints on $\Omega_{\rm m}$ and $\sigma_8$ from 4x2pt+N data vectors. Contours show $68\%$ and $95\%$ uncertainties. Blue contours show the blinded constraints of the fiducial analysis, orange contours show the constraints when allowing the selection biases to vary with redshift, and pink contours show the constraints from a data vector measured without applying systematic weights on \redmapper\ clusters.}
\label{fig:systematicweightchain}
\end{figure}

Fig.~\ref{fig:systematicweightchain} summarizes the tests performed before unblinding the analysis. First, \cite{Simpaper} find that the selection bias ($b_{\rm{sel}}$) in $3$ out of $11$ simulations exhibit redshift evolution at 2 to 3$\sigma$ significance. We test whether adopting a redshift dependent $b_{\rm{sel}}$ model leads to a shift in cosmological constraints.  Second, \redmapper\ clusters might be affected by local survey systematics, although we expect such effects to be much smaller than in the case of galaxies. This is because galaxy clusters are extended objects on the sky, and the random points of \redmapper{} clusters are generated by injecting fake clusters on the sky \citep{redmappersv}, thereby capturing some systematics, such as survey depth variations. To verify this expectation, we follow the method described in \cite{Y1galaxyclustering} to derive a systematic weight for each \redmapper{} cluster. We then analyze the data vector with and without systematic weights and find no changes in the cosmological constraints. Fig.~\ref{fig:systematicweightchain} shows that none of the aforementioned systematics can substantially affect the cosmological constraints. 

\subsection{Cosmological and Nuisance parameters}
\begin{figure}[ht]
\centering
\includegraphics[width=1.0\textwidth]{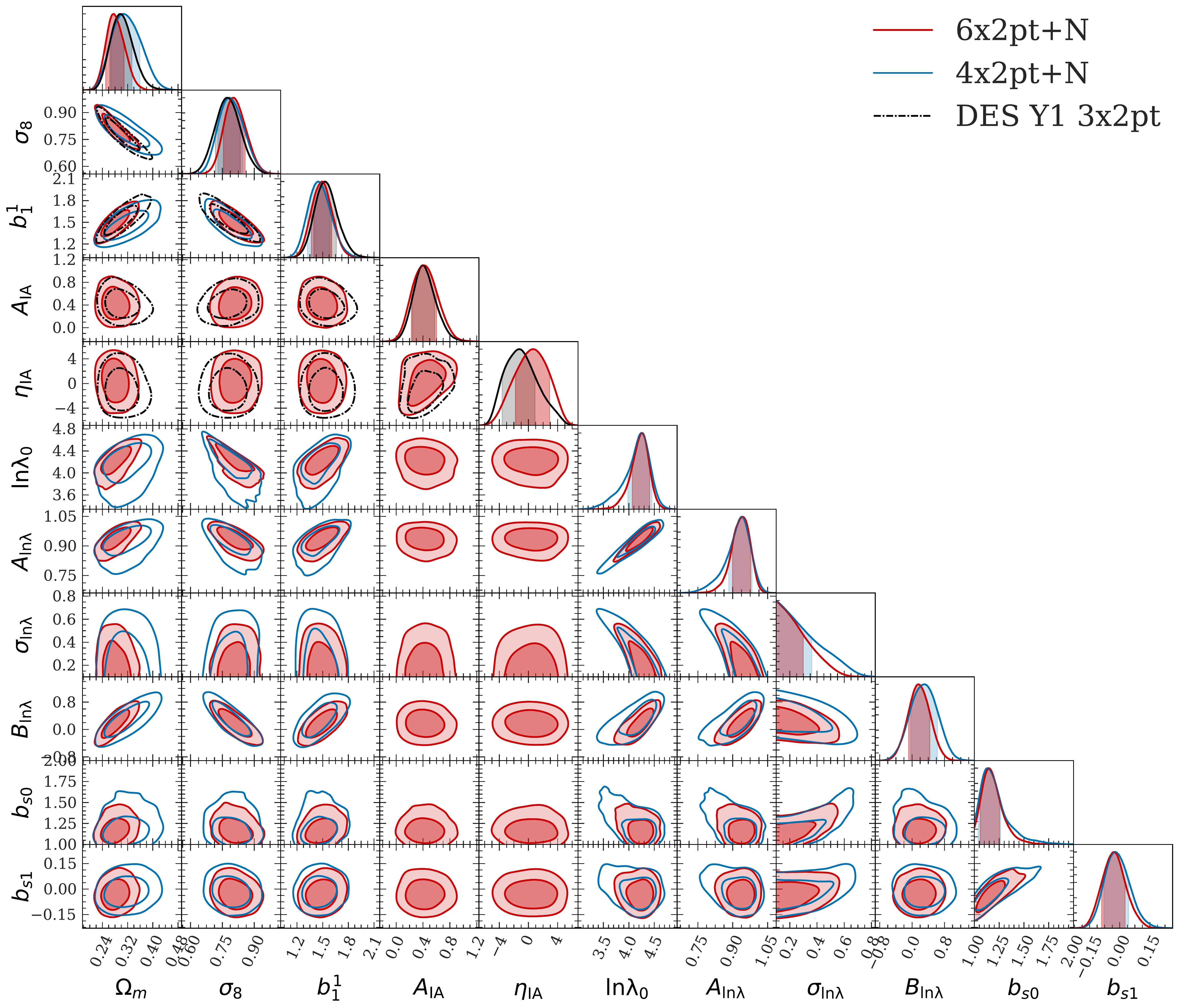}\hspace{-0.05\textwidth}
\caption{Marginalized posterior distributions of the parameters that are not dominated by priors; contours show $68\%$ and $95\%$ confidence levels. We only show galaxy bias for the first redshift bin, since all the bins have qualitatively similar behavior. The constraints are shown for three combinations of the correlation functions and cluster abundances: \ttt\ (black), \clustercomb\ (blue), and \allcomb\ (red).}
\label{fig:allparam}
\end{figure}

The priors and posteriors of the nuisance parameters are summarized in Tab.~\ref{tab:paramsum}. The two dimensional-marginalized posteriors for parameters that are not dominated by priors are shown in Fig.~\ref{fig:allparam}.

\begin{table}
    \centering
    \footnotesize
    \caption{Nuisance parameters considered in the \clustercomb. Parameters labeled with * are additional parameters for the \allcomb\ analysis.  Flat represents the flat prior in the given range and $\rm{Gauss}(\mu, \sigma)$ denotes the Gaussian prior with mean $\mu$ and  width $\sigma$. The third column summarizes the constraints from \clustercomb, and the fourth column summarizes the constraints from \allcomb. The numbers show 1D peaks of the posterior and asymmetric $68\%$ confidence intervals.}
    \label{tab:paramsum}
   \begin{tabular}{lccccc}
    \hline 
	Parameter	 & Prior& \clustercomb\ &\allcomb\ \\
	\hline 
	\hline 
	&Galaxy Bias& \\
	$b_1^1$ &  Flat (0.8, 3.0) & $1.45\pm 0.13$ &$1.49^{+0.11}_{-0.13}  $\\ 
	$b_1^2$ &  Flat (0.8, 3.0) & $1.73\pm 0.13$ & $1.71^{+0.13}_{-0.11}          $ \\ 
	$b_1^3$ &  Flat (0.8, 3.0) &$1.70^{+0.14}_{-0.12}      $& $1.70\pm 0.12        $\\ 
	*$b_1^4$ &  Flat (0.8, 3.0) & -& $2.06^{+0.11}_{-0.16}   $\\ 
	*$b_1^5$ &  Flat (0.8, 3.0) & -&$2.12^{+0.14}_{-0.16}  $\\ 
	\hline
	&\redmagic{} Photo-z&\\
	$\Delta_{z,g}^1$ & $\rm{Gauss}(0.008, 0.007)$ &$0.0080\pm 0.0066          $&$0.0088\pm 0.0066$\\
	$\Delta_{z,g}^2$ & $\rm{Gauss}(-0.005, 0.007)$&$-0.0059^{+0.0060}_{-0.0067}$& $-0.0055^{+0.0073}_{-0.0060}$\\
	$\Delta_{z,g}^3$ & $\rm{Gauss}(0.006,0.006)$&$0.0057^{+0.0056}_{-0.0048}$&$0.0074^{+0.0049}_{-0.0057} $\\
	*$\Delta_{z,g}^3$ & $\rm{Gauss}(0.0,0.01)$&-&$0.0002\pm 0.0091$\\
	*$\Delta_{z,g}^3$ & $\rm{Gauss}(0.0,0.01)$&-& $0.0011^{+0.0092}_{-0.010}  $\\
	\hline
	&Source galaxy Photo-z& \\
	$\Delta_{z,s}^1$& $\rm{Gauss}(-0.001,0.018)$ &$-0.013\pm 0.014           $&$-0.0096^{+0.015}_{-0.013}$\\
	$\Delta_{z,s}^2$ &$\rm{Gauss}(-0.019, 0.013)$ &$-0.021^{+0.010}_{-0.013}  $&$-0.031^{+0.010}_{-0.011}  $\\
	$\Delta_{z,s}^3$& $\rm{Gauss}( 0.009, 0.011)$ &  $0.009\pm 0.010            $& $0.0072^{+0.010}_{-0.0089} $\\
	$\Delta_{z,s}^4$& $\rm{Gauss}(-0.018, 0.022)$ &$-0.019\pm 0.021           $&$-0.018^{+0.016}_{-0.022}  $\\
	\hline
	&shear calibration\\
	$m^1$ & $\rm{Gauss}(0.012, 0.023)$ &  $0.006^{+0.024}_{-0.020}   $&$0.010\pm 0.022 $\\
	$m^2$ & $\rm{Gauss}(0.012, 0.023)$ & $0.009^{+0.024}_{-0.019}   $& $0.012\pm 0.022     $\\
	$m^3$ & $\rm{Gauss}(0.012, 0.023)$ & $0.017\pm 0.021            $&$0.006\pm 0.021     $\\
	$m^4$ & $\rm{Gauss}(0.012, 0.023)$ & $0.012^{+0.022}_{-0.020}   $&$0.014\pm 0.021         $\\
	\hline
	&Intrinsic alignment \\
	*$A_{\rm{IA}}$ & Flat (-5, 5)&- & $0.41\pm 0.19        $\\
	*$\eta_{\rm{IA}}$ & Flat (-5, 5) &-& $0.6\pm 2.3          $\\
	\hline
    &\redmapper{} richness--mass relation \\
    $\rm{ln}\lambda_{0}$ &  Flat (2.0,5.0) &$4.25^{+0.21}_{-0.27}           $&$4.26^{+0.15}_{-0.20}     $\\
    $A_{\rm{ln}\lambda}$ &  Flat (0.1,1.5) & $0.939^{+0.044}_{-0.057}    $&$0.943^{+0.034}_{-0.044}  $\\
    $B_{\rm{ln}\lambda}$ & Flat (-5.0, 5.0) & $0.32^{+0.31}_{-0.35}      $&$0.15^{+0.29}_{-0.24}    $\\
    $\sigma_{\rm{ln}\lambda}$ &  Flat (0.1, 1.0)& $< 0.362  $ &   $< 0.299            $  \\
    \hline 
    &\redmapper{} selection effect \\
    $b_{s0}$ &  Flat (1.0,2.0) &$1.13^{+0.14}_{-0.086}  $&$1.15^{+0.11}_{-0.09}     $\\
    $b_{s1}$ & Flat (-1.0,1.0)&  $-0.023^{+0.068}_{-0.056} $&$-0.029^{+0.056}_{-0.062}$\\
   \hline \vspace{-3mm}\\   
    \end{tabular}
\end{table}
\subsection{Covariance matrix}
\begin{figure}[ht]
\centering
\includegraphics[width=1.0\textwidth]{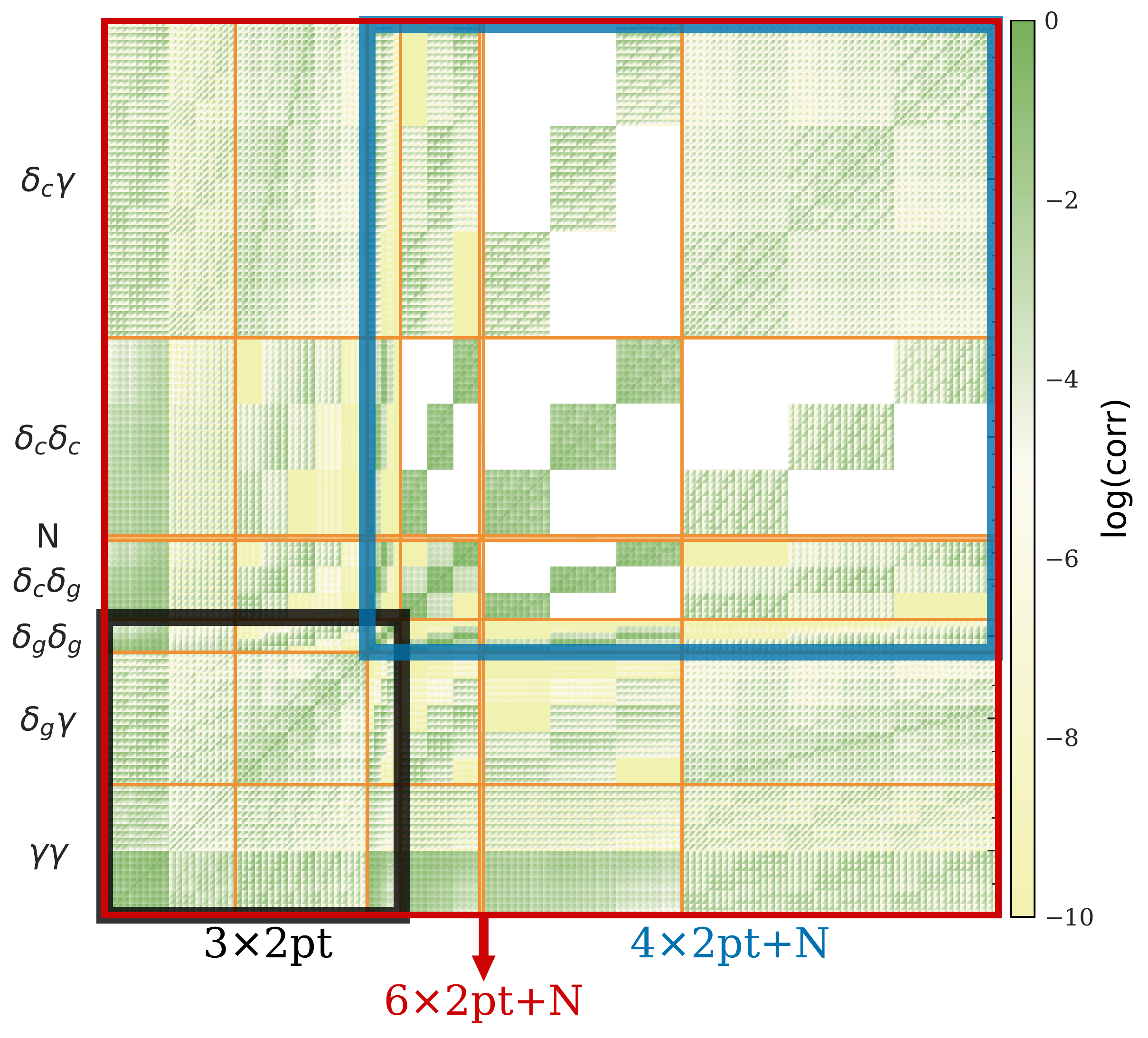}\hspace{-0.05\textwidth}
\caption{Multi-probe correlation matrix for the data vector in this analysis: cosmic shear, galaxy--galaxy lensing, galaxy clustering, cluster--galaxy cross correlations, cluster abundances, cluster--clustering, and cluster lensing. Evidently, different probes are correlated.}
\label{fig:covariance matrix}
\end{figure}

The covariance matrix employed in this analysis is calculated from an analytic model. In brief, the covariance matrix can be separated into three components: the covariance of angular two-point correlations with angular two-point correlations, of cluster abundances with angular two-point correlations, and of the cluster abundances with cluster abundances. %
The first component is described in \citep{DESmethod,Simpaper} and 
modeling of the latter two components is described in \citep{Simpaper}. In Fig.~\ref{fig:covariance matrix}, we show the correlation matrix of cluster abundances and all six two-point correlation functions in this analysis. 

\section{Derivation of cluster mean masses--richness relation}
The mean mass at a given richness ($\lambda$) and redshift ($z=0.35$) can be calculated by 
\begin{equation}
\label{eq:meanmass}
    \langle M|\lambda,z=0.35 \rangle = \frac{\int_0^\infty dM\  M n(M)P(\lambda|M,z=0.35)}{\int_0^\infty dM\ n(M)P(\lambda|M,z=0.35)},
\end{equation}
where n(M) is the halo mass function, and $P(\lambda|M,z)$ is the richness--mass relation. To properly marginalize over cosmological and nuisance parameters, we evaluate equation \ref{eq:meanmass} on a grid of richness from 20 to 120 at each point of the MCMC chain. We then fit a power-law model defined as 
\begin{equation}
    \langle M|\lambda,z=0.35 \rangle = A(\frac{\lambda}{40})^B,
\end{equation}
where A and B are two free parameters, to obtain the mean masses--richness scaling relation.

\end{document}